\documentclass{article}
\usepackage{spconf,amsmath,graphicx,amssymb,hyperref,xurl}
\usepackage[pagewise,modulo]{lineno}
\usepackage[labelsep=period]{caption}

\title{Reverb Conversion of Mixed Vocal Tracks Using an End-to-End Convolutional Deep Neural Network}

\name{$^*$Junghyun Koo$^1$, $^*$Seungryeol Paik$^1$\thanks{*Equal contribution}, Kyogu Lee$^{1,2}$}
\address{
$^1$Music and Audio Research Group, Department of Intelligence and Information,
\\$^2$Artificial Intelligence Institute, Seoul National University\\\{dg22302, paik402, kglee\}@snu.ac.kr}

\begin{document}
\ninept
\maketitle
\begin{abstract}
Reverb plays a critical role in music production, where it provides listeners with spatial realization, timbre, and texture of the music. Yet, it is challenging to reproduce the musical reverb of a reference music track even by skilled engineers.
In response, we propose an end-to-end system capable of switching the musical reverb factor of two different mixed vocal tracks. This method enables us to apply the reverb of the reference track to the source track to which the effect is desired. Further, our model can perform de-reverberation when the reference track is used as a dry vocal source.
The proposed model is trained in combination with an adversarial objective, which makes it possible to handle high-resolution audio samples.
The perceptual evaluation confirmed that the proposed model can convert the reverb factor with the preferred rate of 64.8\%.
To the best of our knowledge, this is the first attempt to apply deep neural networks to converting music reverb of vocal tracks.
\end{abstract}

\section{Introduction}
\label{sec:intro}
Reverberation or reverb is a combination of multiple sound reflections and diffractions that provides spatial perception. In the music industry, reverb is arguably an integral part of most chart-topping music. Generation and control of musical space with reverb is essential because of space realization, timbre and texture modification, and sound cohesion. Artificial reverb systems were developed to overcome the limitations of natural reverb that has a high dependence on initial space \cite{valimaki2012fifty}. Through a digital reverb system based on mathematical algorithms, the software reverb plug-in system is now being generally used along with DAW (Digital Audio Workstation) \cite{leider2004digital}. 

In music production, musicians and engineers often face difficulties of figuring out reverb from reference track and applying it to their own ones. Behind these difficulties, there are three possible reasons as follows.
\begin{enumerate}
    \item Reverb that one listened to may not be from a single reverb plug-in, as use of reverb bus channel or reverb bus, which is separate external channel where reverb plug-in is placed.
    \item Even if reverb plug-ins are detected, it is almost impossible to figure out all subdivided parameter settings inside.
    \item Audio effect units such as equalization and compressor, are often placed at front and back of reverb plug-in which change the sound of the reverb. 
\end{enumerate}
Thus, simply by listening, even such experienced audio engineers cannot easily judge what the applied reverb is and its detailed settings \cite{de2017perceptual}.

Most reverb plug-ins in modern music production are a ‘black-box’ system. These systems generate reverb factors through complex operations, making it difficult for algorithms such as estimating impulse response (IR) \cite{kinoshita2013reverb,larsen2003acoustic} and modeling reverberator \cite{ramirez2020modeling} to infer them. 
To this end, we propose an end-to-end reverb conversion system using a modified version of the U-Net \cite{ronneberger2015u}, which its task is to interchange the musical reverb factors of two different mixed tracks.
As it is the clearest part and placed at the very front of a song, we used stereo-channeled and 44.1kHz of high-fidelity vocal tracks for producing realistic musical reverb.
Our model is not only capable of directly converting the reverb of input audios in a single step but is also capable of de-reverberation.
Through objective evaluation, we show that our model successfully converts the reverb factor of the input and maintains its quality at the same time.
Furthermore, we subjectively evaluated our model and achieved a preferred rate of 64.8\%, despite the difficulty of measuring spatial difference between sounds quantitatively.

\urlstyle{rm}
Examples of the generated audio samples are available at \url{https://dg22302.github.io/MusicalReverbConversion/}.

\begin{figure*}[t]
  \centering
  \includegraphics[width=\linewidth]{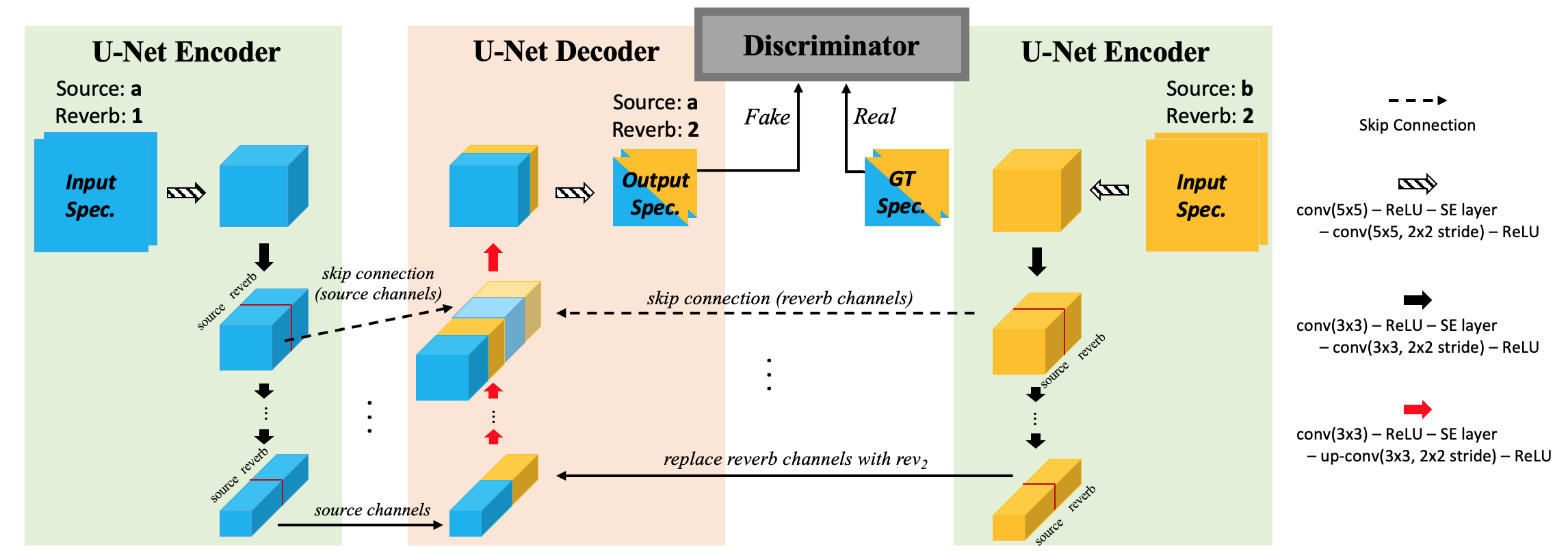}
  \setlength{\abovecaptionskip}{-5pt}
  \setlength{\belowcaptionskip}{-5pt}
  \caption{Overview of the proposed method. The proposed model is a modified version of the U-Net, which is trained to disentangle the reverb factor of the input and convert them into those of counterpart input.}
  \label{fig:model_architecture}
\end{figure*}

\section{Related Works}
\label{sec:related_works}
Deep learning architectures for audio effects have been researched lately for linear effects such as equalization \cite{ramirez2018end, pepe2020designing} and nonlinear effects such as compressor \cite{hawley2019signaltrain}. In the case of the time-varying effect reverb, speech recognition in reverberant environments \cite{delcroix2015strategies} and speech de-reverberation have become a heavily researched field \cite{kothapally2020skipconvnet,wu2016reverberation}. In addition, Deep Neural Networks (DNN) for artificial reverb, commonly used in music production, has been explored in the following areas: generating artificial reverb by IR estimation \cite{ly2020generating}, emulating and modeling of reverb algorithm \cite{ramirez2020modeling, de2017ten}, parameter automation, and intelligent control of reverb plug-in \cite{chourdakis2016automatic}, and matching speech spaces by adding artificial reverb \cite{sarroff2020blind}.

Prior to our work, end-to-end DNN for extracting and applying reverb from musically mixed vocals has not yet been proposed. Due to the three reasons in the introduction - the presence of multiple reverb bus channels, variously subdivided parameters inside the plug-ins and the influence of other audio effect plug-ins in the channels - and the black-box algorithms of plug-ins, replicating the musical reverb of the reference track is nearly impossible to be achieved with previous methods.
Therefore, to solve the problem, we use a modified version of the U-Net to disjoint musical reverb factor from the input mixture signal of a high resolution.

\section{Proposed Method}
\label{sec:proposed_method}
U-Net, the model originally tasked to perform a fast and precise segmentation of images, had also shown prodigious performance in speech related subjects, such as speech enhancement \cite{choi2018phase}, and source separation \cite{jansson2017singing}. It is excellent with both modifying and preserving the network inputs, where skip-connections have shown to play a crucial role.
In our case, retaining the contents of the input audio is significant for replacing the reverb factor.
For that reason, we propose a modified version of the U-Net that disjoints the source and reverb factor of the input.
This idea of disentanglement using a neural network was inspired by the method proposed in \cite{zhang2017split}.

\subsection{Input representation}
\label{sec:data}
For input data to train our model, reverb mixed tracks were used, based on the understanding of reverb application in mixing music. In the procedure, a source track, which is dry recorded sound, is needed. Source track signal is sent to reverb bus channel, where `Wet 100\%' parameter setting reverb plug-in is placed. The output signal of the channel is 100\% reverberated source signal and it is combined with the source signal by a certain percentage, called bus send ratio. 

Following the process, we mixed dry source tracks with their corresponding fully reverberated tracks at a certain ratio, which works as reverb bus, for generating model inputs :
\begin{equation}
    \cfrac{1}{\gamma + 1} ({{src}+{\gamma \cdot rev}}).
\label{equation:input2}
\end{equation}
The reverb mixed track can be represented as a sum of source signal $src$ and 100\% reverberated source signal $rev$, with a factor of bus send ratio $\gamma$. Dividing by ($\gamma$ + 1), prevention of possible peak occurs and normalization can be achieved. For convenience, let us abbreviate equation (\ref{equation:input2}) as $s_x r_\#$, a mixture of source $x$ and reverb factor $\#$. For instance, a model input of source $a$ combined with reverb $1$ is denoted as $s_a r_1$.

\subsection{Model architecture}
\label{model}
As illustrated in Figure~\ref{fig:model_architecture}, the proposed network is a fully convolutional network, aimed to convert the reverb factor of two input spectrograms that are combinations of $s_a r_1$, and $s_b r_2$.
The converting procedure is done with attempts to disentangle the source and reverb factor of outputs from each layer by channels, where the first half channels is that of source, and the other half is that of reverb.
The disentangle procedure is carried out by converting reverb channels of the outputs at the encoder's last layer, and by channel-wise concatenating each reverb channels from other encoding layers to the counterpart decoding layers.
Therefore, in the case where the input is $s_a r_1$, the U-Net decoder is intended to decode the source factor of the input spectrogram, and the reverb factor of $s_b r_2$ to generate output of $s_a r_2$.
Lastly, the discriminator takes the output spectrogram to train the network in an adversarial manner.

Additional modifications have been made, including the method explained above, as follows. First, each layer of the encoder and decoder is a form of a \textit{Squeeze-and-Excitation} (SE) \cite{hu2018squeeze} block, where each block consists of two convolutional layers with a SE layer in between them. The kernel size for each convolutional layer is 3x3 except for the uppermost layer, where the kernel size is 5x5. Each convolutional layer is followed by a rectified linear unit (ReLU) activation. Second, instead of encoding the feature dimension with max-pooling, we use a convolutional stride factor of 2x2 at the second convolutional layer in each block. Similarly, for decoder, transposed convolution with the same stride factor is performed to up-sample feature dimension. Finally, skip connection is not performed at the first layer.

\subsection{Training}
\label{training}
As our model requires two inputs ($s_a r_1$ and $s_b r_2$) to convert the reverb factor, we train our model by calculating both losses for each result simultaneously. For convenience, we explain our loss term in the case where the network converts input $s_a r_1$ to $s_a r_2$.

Our training loss is a combination of a spectrogram term, latent term, and an adversarial term.
The spectrogram loss $\mathcal{L}_{spec}$ penalizes the distance between output of the U-Net's decoder and ground truth with mean absolute error and mean squared error. By assuming that our network can successfully disjoint reverb and source factors by channels, we add a latent loss $\mathcal{L}_{latent}$ to penalize the distance between reverb channels generated from each layer of the encoder.
For $f_1 \circ f_2 \circ \cdots f_n = g_n$, where $f_n$ is $n^{\text{th}}$ layer of the encoder, the loss term $\mathcal{L}_{latent}$ is as follow:
\begin{equation}
\begin{split}
    \mathcal{L}_{latent} = \sum_{i=1}^{n} ( [[g_i(s_a r_1)]_{rev} -[g_i(s_b r_1)]_{rev}| \\
    + \|[g_i(s_a r_1)]_{rev} -[g_i(s_b r_1)]_{rev}\|_2^2 ).
\end{split}
\end{equation}
Here, $[g_n]_{rev}$ corresponds to reverb channels of the $n^{\text{th}}$ layer output. Since $\mathcal{L}_{latent}$ only penalizes the reverb channels, we use two different sources with the same reverb as the encoder's inputs.

With the expectation of generating a realistic reverberant sound, an adversarial loss $\mathcal{L}_{adv}$ is used, which follows the original GAN loss introduced in \cite{goodfellow2014generative}. The adversarial loss terms composed of the discriminator loss $\mathcal{L}_{D}$ and the generator loss $\mathcal{L}_{\psi}$ are as follows,
\begin{equation}
\begin{split}
    \mathcal{L}_{D} &= \mathbb{E}_{s_a r_2}[log(D(s_a r_2))] \\
    &+ \mathbb{E}_{s_a r_1, s_b r_2}[log(1-D(\psi(s_a r_1, s_b r_2)))], \\
    \mathcal{L}_{\psi} &= \mathbb{E}_{s_a r_1, s_b r_2}[log(1-D(\psi(s_a r_1, s_b r_2)))],
\end{split}
\end{equation}
where $\psi$ represents the proposed model's encoder and decoder combined. The architecture of the discriminator $D$ is almost equivalent to the encoder, but an additional convolutional layer was used at the end. The kernel size of the last layer is 4x3, where the input was max pooled to match this shape. The final probability is produced with a sigmoid activation function.
To sum up, $\mathcal{L}_{D}$ is maximized, where $\mathcal{L}_{spec}$, $\mathcal{L}_{latent}$, and  $\mathcal{L}_{\psi}$ are minimized over the training process.

\section{Experiments}
\label{sec:experiments}

\subsection{Dataset}
\label{ssec:dataset}
Our dataset is composed of stereo channeled dry and reverberated vocal files, where the sampling rate is 44.1kHz and the bit rate 16-bits. 
The total duration of the dataset is 32 hours and 10 minutes of clean studio-recorded dry vocal tracks - singing, rapping, and speaking voices by thirty-four multilingual vocalists, but mainly Korean. We split the dataset into 30 and 2 hours for the training and validation set, respectively. From the clean source, corresponding fully reverberated tracks are generated through multiple reverb plug-ins.

For the train set, we randomly set $\gamma$ of ratio between 0\% and 75\% with 5\% intervals to generate network inputs. Thirty-six reverb presets from three Valhalla DSP reverb plug-ins – ValhallaVintageVerb, ValhallaPlate and ValhallaRoom - were used to generate training data whose samples were randomly chosen during training. For the validation set, four reverb presets from three reverb plug-ins - ChromaVerb of Logic Pro X, H-Reverb and Abbey Road Plates of Waves - were used to generate data with various combinations of presets and $\gamma$.
All our evaluations shared fixed samples that could be obtained from the validation set.

\subsection{Experimental setups}
\label{ssec:exp_setup}
Regarding the implementation of the proposed model, a Hamming window of 2048 samples with 75\% overlap, and 2048-point fast Fourier transform (FFT) were employed for the STFT analysis. Only the magnitude spectrograms were used as network inputs, where the final audio is computed with the network output's magnitude and the input's phase. The length of input audio is 7.43 seconds, and the number of steps per epoch is the total length of training data divided by the input audio length.

The encoder and decoder of the proposed model consist of 5 layers, where the number of channels increases to 32, 64, 128, 256, 512 as the input is encoded.
During the training stage, ground truth spectrograms were also encoded through the encoder to compute $\mathcal{L}_{latent}$. The $\mathcal{L}_{adv}$ was applied after 20 epochs of training, and the training was stopped at 100 epoch. We use Rectified Adam optimizer \cite{liu2019variance} with a learning rate of 0.001 and momentum parameters $\beta$1 = 0.9, $\beta$2 = 0.999.

\begin{table}[t]
\footnotesize
\setlength{\belowcaptionskip}{-15pt}
\caption{Model description.}
\centering
\begin{tabular}{|c|c|}
\hline
\textbf{Model} & \textbf{Method}                                                                                              \\ \hline
model 1        & \begin{tabular}[c]{@{}c@{}}w/o (skip-connect reverb channels\\ \& discriminator \& latent loss)\end{tabular} \\ \hline
model 2        & w/o (discriminator \& latent loss)                                                                           \\ \hline
model 3        & w/o (latent loss)                                                                                            \\ \hline
model 4        & proposed                                                                                                     \\ \hline
\end{tabular}
\vspace{-10pt}
\label{table:model_description}
\end{table}

\subsection{Quantitative evaluation}
\label{ssec:quant_eval}
The quantitative evaluation includes two tasks; \textit{Reverb Conversion}: interchanging reverb of two different inputs and \textit{De-reverberation}: eliminating reverb of the target input.
Four different proposed models with specifications indicated in Table \ref{table:model_description} were compared for both evaluations, where a WPE-based system \cite{nakatani2010speech} was included as the baseline model in the de-reverberation task. We measure \textit{Reverb Conversion} with the wideband measurement of the perceptual evaluation of speech quality (PESQ) \cite{rec2005p} and short-time objective intelligibility (STOI) \cite{taal2010short}, which are metrics frequently used in tasks regarding speech enhancement. Speech-to-reverberation modulation energy ratio (SRMR) \cite{santos2014improved} and scale-invariant source-to-noise ratio (SI-SDR) \cite{le2019sdr} are further measured for \textit{de-reverberation}.
In the case of PESQ, the samples are down-sampled to 16kHz for computation.
A higher value represents a better result in all the metrics used.

\begin{figure*}[t]
\begin{minipage}[b]{0.24\linewidth}
  \centering
  \centerline{\includegraphics[width=4.5cm]{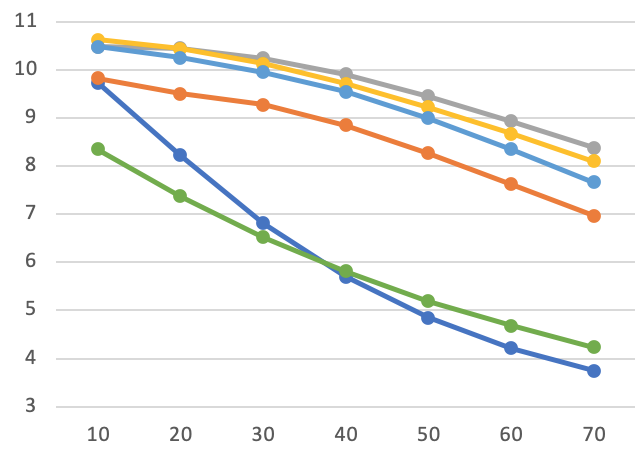}}
  \centerline{(a) SRMR}\medskip
\end{minipage}
\begin{minipage}[b]{0.24\linewidth}
  \centering
  \centerline{\includegraphics[width=4.5cm]{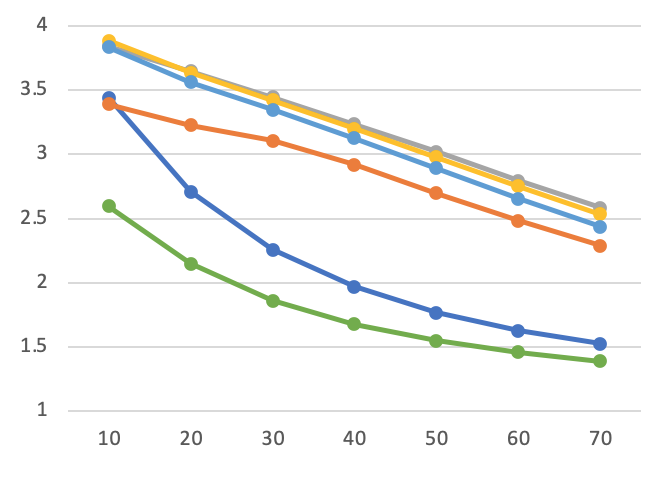}}
  \centerline{(b) PESQ}\medskip
\end{minipage}
\begin{minipage}[b]{0.24\linewidth}
  \centering
  \centerline{\includegraphics[width=4.5cm]{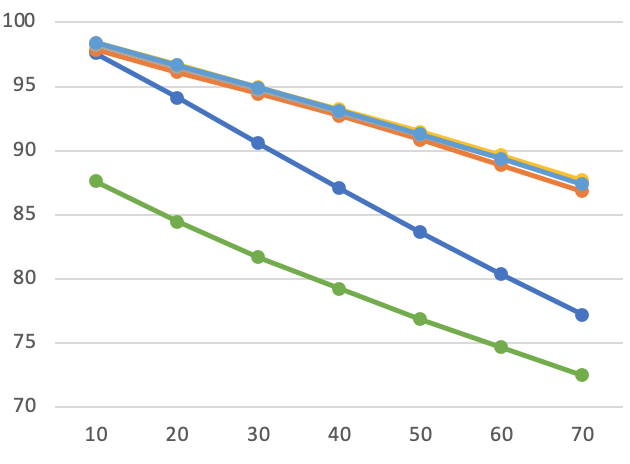}}
  \centerline{(c) STOI}\medskip
\end{minipage}
\begin{minipage}[b]{0.24\linewidth}
  \centering
  \centerline{\includegraphics[width=4.5cm]{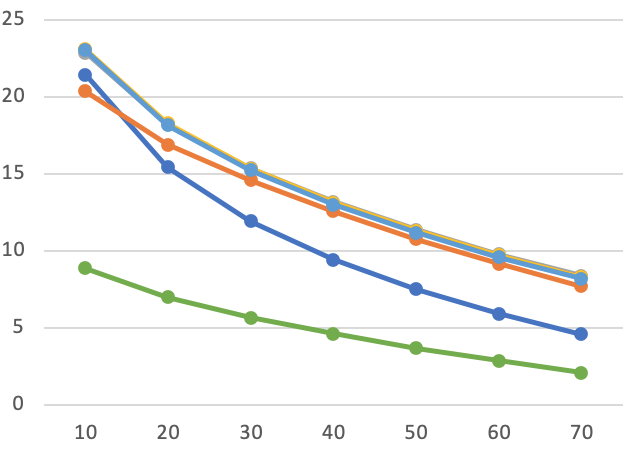}}
  \centerline{(d) SI-SDR}\medskip
\end{minipage}
\begin{minipage}[t]{0.01\linewidth}
\vspace{-24\linewidth}
  \centering
  \centerline{\includegraphics[width=1.1cm]{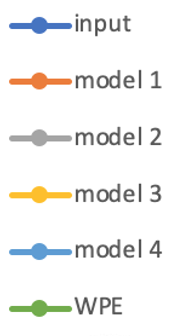}}
\end{minipage}
\caption{Results of de-reverberation with four different quantitative metrics. Values on the x-axis are the percentage of bus send ratio ($\gamma$) set for mixing source and reverb factor. The unit for STOI is percent (\%), and SRMR and SI-SDR are in decibel (dB).}
\label{fig:derev_results}
\end{figure*}

\begin{table}[t]
\footnotesize
\caption{Results of reverb conversion.}
\centering
\begin{tabular}{c|c|c}
\hline
\textbf{Methods}  & \textbf{PESQ}  & \textbf{STOI (\%)} \\ \hline
input             & 2.643          & 78.00              \\ \hline
model 1           & 3.026          & 80.08              \\
model 2           & 3.099          & 80.56              \\
model 3           & 3.098          & \textbf{80.92}     \\
model 4           & \textbf{3.105} & 80.77              \\ \hline
\end{tabular}
\vspace{-10pt}
\label{table:rev_conv}
\end{table}

\subsubsection{Reverb Conversion}
\label{ssec:rev_conv}
Conventionally, metrics used in speech enhancement are a comparison between clean and enhanced samples, where most metrics are weighted more on the contents rather than on other acoustic effects, including reverb.
However, since our task is to convert the reverb factor from one to another, we aim to measure the difference, including the reverb factor. Consequently, we evaluate our metrics with a comparison between target reverberated and interchanged samples to differentiate from speech enhancement. For this, we measure the results with PESQ and STOI, which are metrics able to convey the reverb factor through prior experiments.

We solely compare our models with the input since there are no other models available for this specific task. From Table \ref{table:rev_conv}, it is observed that the evaluated values of the input were relatively high; yet, all of our models’ evaluated scores surpassed that of the input.
Noting that the input shares the identical source of the target’s, we can intuitively realize that our models not only could maintain the contents of the input but also successfully converted the reverb factor to that of the target.
Furthermore, models using the discriminator were better at converting the reverb factor than models that did not.
We later qualitatively evaluate \textit{Reverb Conversion} in subsection \ref{ssec:listening_test} to supplement contents that could not be addressed with objective measure.

\subsubsection{De-reverberation}
\label{ssec:derev}
Although our main objective is not de-reverberation, evaluating this task could objectively demonstrate the performance of our models along with the baseline model.
Given two inputs $s_a r_1$ and $s_b r_2$, our network can perform de-reverberation of $s_a r_1$ when $\gamma_2$ is set to 0. With this example, Fig.\ref{fig:derev_results} shows the model results measured according to the ratio of $\gamma_1$.
Note that even when $\gamma_1$ is set to 70\%, the actual source-to-reverb ratio is 59\% to 41\%, using equation \eqref{equation:input2}.

All of our models showed an eased decay for the increasing percent of $\gamma$ in all metric scores than the input and baseline model.
The WPE achieved lower scores than the input with evaluations except for SRMR, a non-intrusive metric. From here, scores of the baseline model start to exceed input's after $\gamma$ of 40\%.
Given that audio contents is the main factor in the evaluations, our result can be interpreted as not because the WPE system was not capable of de-reverberation, but that there was a content degradation for the baseline system while the input retained the high quality of the clean source.
Nevertheless, the performance of our model further exceeded both input and the baseline as the $\gamma$ increased, while there was little difference in performance among model 2, 3, and 4.
From this, we can infer that in de-reverberation task, our model successfully maintained both the contents and quality of the input regarding its high resolution.

\begin{table}[t]
\footnotesize
\caption{Preferred rate of GT}
\centering
\label{tab:gt_pref}
\begin{tabular}{c|c|c|c}
\hline
\textbf{Status} & \textbf{Overall (\%)} & \textbf{W→D (\%)}    & \textbf{D→W (\%)} \\ \hline
Total           & 64.8            & 85.7          & 43.9       \\ \hline
Experienced     & 62.1            & 81.7          & 43.3       \\ \hline
Inexperienced   & 67.5            & 89.6          & 44.4       \\ \hline
\end{tabular}
\end{table}

\subsection{Listening test}
\label{ssec:listening_test}
We conducted listening test with twenty participants. Ten participants were musicians and sound engineers denoted as experienced, and the other ten participants denoted as inexperienced were not familiar with critical listening. 
The participants were randomly given one of two different test sets with twenty-four questions each. The questions were equally distributed by considering presets and $\gamma$ differences between two inputs ($\Delta\gamma$), which were set at 0\%, 20\%, 40\% and 60\%.
Among the chosen samples, there were no samples with $\gamma$ that exceeded 75\%, and samples of 0\% $\gamma$ when $\Delta\gamma$ was set to 0\%.
For each question, three samples are presented - a reference sample, which is the output of the proposed model, with two different samples, which are an input of the model and ground truth of the reference sample (GT).
The aim of the test was to assess the performance of our model's reverb conversion, by identifying whether the sound of GT is preferred to have more similar spatiality - a sense of space - with reference sound of model output, compared to that of input in actual listening.

As shown in Table \ref{tab:gt_pref}, experienced participants were less likely to feel similar spatiality between reference and GT sounds than inexperienced.
Because the presented samples share identical vocal segments of high-resolution, it was hard to assess the spatial difference with presets and $\Delta\gamma$.
Thus, we listened then manually distinguished the input and GT samples of each question with a drier or wetter reverb factor, except for six ambiguous questions.

Based on the analysis, we approached preference for GT by dividing generated samples into two groups: W→D and D→W. W→D represents samples generated with input’s wetter sound converted to target’s drier sound. For example, suppose the model receives an input track $s_a r_1$ and a reference track $s_b r_2$. When $r_1$ has a wetter reverb factor than $r_2$, the model’s output is considered as W→D. Contrariwise, D→W represents samples generated with drier $r_1$ and wetter $r_2$.
Regarding the preference tendency, preferred rate of W→D was higher than D→W across all participants. This perceptual finding confirms that the proposed model especially shows better conversion performance on drier $r$ application, similar to the de-reverberation process, than the conversion performance on wetter $r$ application. This can be reasoned with our network's limitation of generating only the magnitude of the input spectrogram, where the phase is not inferred. To be specific, in the case of D→W, the output magnitude containing the generated reverb factor will be converged with the input phase and create an artificial noise, which may be unpleasant to listeners.

\section{Conclusion}
\label{sec:conclusion}
The proposed model is an end-to-end system that converts the musical reverb factor of the mixed vocal track into that of the reference track with the fully convolutional architecture.
We handle audio with high-resolution, and our network generated samples that preserved the quality and the contents of input.
The evaluation showed that despite this being a challenging task, our model could successfully interchange the reverb factor from given inputs.
Our model is also capable of de-reverberation and showed promising results compared to the baseline.

A couple of points can be noted as future work. First, phase inference and different models can be applied for generating a more realistic reverb. Second, by assuming that various instrumental tracks are available, our model has the potential of converting the reverb of both single-track and multi-tracks of instruments.



\vfill\pagebreak

\bibliographystyle{IEEEbib}

\bibliography{refs.bib}

\end{document}